\documentclass[preprint]{article}
\newcommand{\be}{\begin{equation}} \newcommand{\ee}{\end{equation}} 
\newcommand{\bea}{\begin{eqnarray}}\newcommand{\eea}{\end{eqnarray}}

\begin{document}
\title{ Deconstructing non-dissipative non-Dirac-hermitian relativistic
quantum systems} 
\author{ Pijush K. Ghosh}
\date{Department of Physics, Siksha-Bhavana,\\ 
Visva-Bharati University,\\
Santiniketan, PIN 731 235, India.}
\maketitle
\begin{abstract} 
A method to construct non-dissipative non-Dirac-hermitian relativistic quantum
system that is isospectral with a Dirac-hermitian Hamiltonian is
presented. The general technique involves a realization of the basic
canonical (anti-)commutation relations involving the Dirac matrices and
the bosonic degrees of freedom in terms of non-Dirac-hermitian operators,
which are hermitian in a Hilbert space that is endowed with a pre-determined
positive-definite metric. Several examples of exactly solvable
non-dissipative non-Dirac-hermitian relativistic quantum systems are presented
by establishing/employing a connection between Dirac equation and supersymmetry.
\end{abstract}
\tableofcontents{}

\section{Introduction}

An emergent view in the current literature is that a formulation of
non-dissipative quantum systems is possible without the requirement
of the Dirac-hermiticity\cite{bend,ali,quasi}. The quantum systems constructed
in this way are indeed hermitian in a Hilbert space that is endowed with a
positive-definite metric different from the identity operator and hence, a
modified inner-product. The construction of the metric in the Hilbert space
is a non-trivial task
and is not known for a majority of the examples considered in this context.
A lack of knowledge of the metric makes even an otherwise exactly solvable
non-Dirac-hermitian quantum system incomplete, since no expectation values
of physical variables can be calculated explicitly. An approach taken in
Refs. \cite{piju_lat,piju1} in this regard was to construct non-dissipative
non-Dirac-hermitian quantum systems with a pre-determined positive-definite
metric in the Hilbert space. The quantum systems constructed following this
approach may serve as prototype examples of non-dissipative non-Dirac-hermitian
quantum models with a complete description. These systems may be used to
test different ideas related to the subject\cite{piju_lat,piju1,me},
including validity of any approximate and/or numerical method.

Recently, non-Dirac-hermitian relativistic quantum systems with entirely
real spectra have been studied\cite{omar,anjana,sever,dutra,cana,bhaba,longhi}.
An optical realization of relativistic quantum system with non-Dirac-hermitian
interaction has also been proposed\cite{longhi}. This proposal assumes a
significance within the background that ideas emanating from combined parity
and time-reversal symmetric optical systems have been verified
experimentally\cite{expt1,expt2}. A majority of the non-Dirac-hermitian
relativistic models proposed so far are incomplete in the sense that the
positive-definite metric and/or the modified inner-product in the Hilbert
space is not known.

The purpose of this article is to present a class of non-dissipative
non-Dirac-hermitian relativistic quantum systems admitting entirely real
spectra and unitary time evolution. The metric and hence, the inner-product
in the Hilbert space is specified explicitly so as to facilitate the
computations of expectation values of different physical observables.
The general method is an extension of the technique employed in Refs.
\cite{piju_lat,piju1} to relativistic systems. In particular, a
non-Dirac-hermitian realization of Dirac matrices and the bosonic variables
is used, which are hermitian in a Hilbert space that is endowed with a
pre-determined positive-definite metric. Apart from the general results,
several examples of exactly solvable non-dissipative non-Dirac-hermitian
quantum systems are presented by employing a connection between supersymmetry
and Dirac equation.

\section{Preliminaries}

The Pauli matrices $\sigma^a$, position variables $x^a$ and the momentum
operators $p^a$ are taken to be hermitian in the Hilbert space ${\cal{H}}_D$
with the standard inner-product $\langle \cdot | \cdot \rangle$. A
Hilbert-space ${{\cal{H}}_{\eta_+}}$ that is endowed with the positive-definite
metric $\eta_+$,
\be
\eta_+ := exp \left ( - 2 \vec{J} \cdot \hat{n} \phi \right ),
\ \ \vec{J} := \vec{L} + \frac{\vec{\sigma}}{2}, \ \
\hat{n} \cdot \hat{n} = 1, \ (\phi, n^a) \in R, \ \ a=1, 2, 3,
\label{metric}
\ee
\noindent and the inner-product
$\langle \langle \cdot | \cdot \rangle \rangle_{{\cal{H}}_{\eta_+}}
:= \langle \cdot | \eta_+ \cdot \rangle$ is introduced. The operator
$\vec{L}$ with $L^a=\epsilon^{abc} x^b p^c$ denotes orbital angular momentum
in ${\cal{H}}_D$, where $\epsilon^{abc}$ denotes the three dimensional
Levi-Civita tensor. The operator $\vec{J}$ can thus be identified
as the total angular momentum of the system. It may be noted that
the metric $\eta_+$ can be decomposed as a direct product of a
purely bosonic and purely fermionic metrics. In particular, 
$\eta_+= \eta_b \otimes \eta_f$ where
$\eta_b:= exp(-2 \vec{L} \cdot \hat{n} \phi)$ 
and $\eta_f:= exp(- \vec{\sigma} \cdot \hat{n} \phi)$. The positivity
of the metric follows from the fact that the eigen values of the operator
$\vec{J} \cdot \hat{n}$ are real. With the introduction of the similarity
operator $\rho$, 
\be
\rho := \sqrt{\eta_+}=exp \left ( - {\vec{J} \cdot \hat{n}}
\phi \right ),
\label{similarity}
\ee
\noindent a set of non-Dirac-hermitian matrices $\Sigma^a$, position variables
$X^a$ and the momentum operators $P^a$ may be introduced as follows\cite{ali,
piju_lat, piju1}:
\bea
\Sigma^a & := & (\rho)^{-1} \sigma^a \rho  =
\sum_{b=1}^3 R^{ab} \sigma^b,\nonumber \\
X^a & := & (\rho)^{-1} x^a \rho  =
\sum_{b=1}^3 R^{ab} x^b,\nonumber \\
P^a & := & (\rho)^{-1} p^a \rho  =
\sum_{b=1}^3 R^{ab} P^b,\nonumber \\
R^{ab} & \equiv & n^a n^b \left ( 1 - cosh \phi \right ) +
\delta^{ab} cosh \phi + i \epsilon^{abc} n^c sinh \phi.
\eea
\noindent The non-Dirac-hermiticity of $\Sigma^a$, $X^a$ and $P^a$ follows
from the fact that $(R^{ab})^* \neq R^{ab}$, where $f^*$ denotes the complex
conjugate of $f$. It can be shown by using the identity
$\sum_{a=1}^3 R^{ab} R^{ac} = \delta^{bc}$ that $\vec{\Sigma} \cdot \vec{P} =
\vec{\sigma} \cdot \vec{p}$. This implies that $\vec{\sigma} \cdot \vec{p}$
and $\vec{\Sigma} \cdot \vec{P}$ are hermitian both in ${\cal{H}}_D$ as well as
in ${\cal{H}}_{\eta_+}$. The matrices $\Sigma^a$ obey the same algebra
satisfied by the Pauli matrices:
\bea
&& \left [ \Sigma^a, \Sigma^b \right ] = 2 i \epsilon^{abc} \Sigma^c, \ \
\left \{ \Sigma^a, \Sigma^b \right \} = 2 \delta^{ab}\nonumber \\
&& \Sigma_{\pm} := \frac{1}{2} \left ( \Sigma^1 \pm i \Sigma^2 \right ), \ \
\{\Sigma_-, \Sigma_+ \} = 1, \ \ \Sigma_{\pm}^2 = 0,
\eea
\noindent and are hermitian in the vector space ${{\cal{H}}_{\eta_+}}$.
Similarly, the position and the momentum operators satisfy the
basic canonical commutation relations,
\be
\left [ X^a, X^b \right ] = 0 = \left [ P^a, P^b \right ], \ \
\left [ X^a, P^b \right ] = i \delta^{ab},
\ee
\noindent and are hermitian in ${{\cal{H}}_{\eta_+}}$. All other commutators
involving $\Sigma^a$, $X^a$ and $P^a$ vanish identically.

\section{Dirac Equation in $1+1$ dimensions}
In this section, a convenient notation $x^1\equiv x$ and $p^1\equiv p_x$ will
be used to denote the position and the momentum operators.
A non-Dirac-hermitian relativistic Hamiltonian in $1+1$ dimensions is
introduced as follows,
\bea
&& H_{1D} = \sigma^2 p_x + \sigma^1 A(x) + \sigma^3 B(x) + V(x)\nonumber \\
&& A(x) \equiv M(x) cosh \phi + i P(x) sinh \phi,\nonumber \\
&& B(x) \equiv i M(x) sinh \phi - P(x) cosh \phi, \phi \in R,
\eea
\noindent where $M(x)$, $P(x)$ and $V(x)$ are real functions of the coordinate.
The complex functions $A(x)$ and $B(x)$ can be interpreted as
scalar and pseudo-scalar potentials, respectively. The potential $V(x)$
is the time-component of a Lorentz two-vector potential. The space-component
of the same two-vector has been taken as zero, since it can always be gauged
away in $1+1$ dimensions. The Lorentz covariance of the Dirac equation can be
shown by introducing $\gamma^0:=\sigma^1$, $\gamma^1:= i \sigma^3$. The matrix
$\gamma_5$ may be defined as $\gamma^5= i \sigma^2$. Various representations
of the Dirac equation in $1+1$ dimensions have been used previously\cite{
anjana,dutra,longhi,khare}. All these representations are related to each other
through unitary transformations. For example, the unitary operator
$U=exp(-\frac{i\pi}{4} \sigma^1)$ relates the representation used in this
article to the one used in Ref. \cite{longhi} in the context of optical
realization of non-Dirac-hermitian relativistic quantum systems.

The Hamiltonian $H_{1D}$ is non-Dirac-hermitian. With the introduction of the
hermitian and positive-definite operators,
\be
\rho_{1D} := e^{- \frac{\phi}{2} \sigma^2}, \ \ 
\eta_{1D}:= \rho_{1D}^2=e^{-\phi \sigma^2},
\ee
\noindent the non-Dirac-hermitian Hamiltonian $H_{1D}$ can be mapped to
a Dirac-hermitian Hamiltonian,
\bea
h_{1D} & = & \rho_{1D} H_{1D} \rho_{1D}^{-1}\nonumber \\
& = & \sigma^2 p_x + \sigma^1 M(x) - \sigma^3 P(x) + V(x).
\eea
\noindent The similarity operator $\rho_{1D}$ is obtained from Eq.
(\ref{similarity}) by choosing $\vec{L}=0, \hat{n}^1=0=\hat{n}^3, \hat{n}^2=1$.
The Hamiltonian $H_{1D}$ is hermitian in the Hilbert space
${\cal{H}}_{\eta_{1D}}$ that is endowed with the metric $\eta_{1D}$ and
a modified norm $\langle \langle \cdot | \cdot \rangle \rangle_{\eta_{1D}}=
\langle \cdot|\eta_{1D} \cdot \rangle$. The Hamiltonian $H_{1D}$ and $h_{1D}$
are isospectral. The function $M(x)$ and $P(x)$ may be identified as the
scalar and the pseudo-scalar interactions, respectively, for the
Dirac-hermitian Hamiltonian $h_{1D}$.

\subsection{Case I: Pure scalar interaction}
The eigen value equation for $h_{1D}$ can be solved exactly for
specific choices of the interactions. The first example considered here is
the vanishing vector potential and vanishing $P(x)$. The Hamiltonian $h_{1D}$
for this case may be identified as the supercharge of a supersymmetric quantum
system with $H_s:=h_{1D}^2$ as the corresponding supersymmetric
Hamiltonian\cite{khare}. In particular,
\be
H_s = p_x^2 + M^2 - \sigma^3 M^{\prime},
\ee
\noindent where $f^{\prime}(x)$ denotes the derivative of $f(x)$ with respect
to its argument. An exhaustive list of $M$ for which $H_s$ is exactly solvable
is known\cite{khare}. The non-Dirac hermitian Hamiltonian $H_{1D}$,
\be
H_{1D} = \sigma^2 p_x + \sigma^1 M(x) cosh \phi + i \sigma^3 M(x) sinh \phi,
\ee
\noindent is thus exactly solvable corresponding to each $M$ for which $H_s$
is exactly solvable. The energy eigen values $E$ of $H_{1D}$ is determined
as $E=\pm \sqrt{E_s}$, where $E_s$ is the energy of $H_s$. It may be
recalled here that $E_s$ being the energy of a supersymmetric theory, it is
always semi-positive definite. The eigen-spinor $\Psi$ of $H_{1D}$ is
related to the eigen functions $\psi$ of $H_s$ as,
$\Psi= \rho_{1D}^{-1} \psi$. A complete set of orthonormal eigen functions
$\psi$ in ${\cal{H}}_{D}$ ensures orthonormality of
$\Psi$ in ${\cal{H}}_{\eta_{1D}}$.

\subsection{Case II: Scalar \& Pseudo-scalar interactions}
The second example describes the case of a vanishing vector potential and
non-vanishing $M(x)$ and $P(x)$. The square of the operator
$h_{1D}$ for this limiting case reads,
\be
h_{1D}^2 = p_x^2 + M^2 + P^2 - \sigma^3 M^{\prime} - \sigma^1 P^{\prime}.
\ee
\noindent A unitary transformation $U:=exp(-i \frac{\theta}{2} \sigma^2),
\theta \in [0, 2 \pi]$ maps $h_{1D}^2$ to the Hamiltonian $H_1$,
\bea
H_1 & := & U^{-1} h_{1D}^2 U\nonumber \\
&=& p_x^2 + M^2 + P^2 + \sigma^1 ( M^{\prime} sin \theta - P^{\prime}
cos \theta )\nonumber \\
& - & \sigma^3 ( P^{\prime} sin \theta + M^{\prime} cos \theta).
\eea
\noindent It may be noted that $U$ is unitary in ${\cal{H}}_D$ as well as in
${\cal{H}}_{\eta_{1D}}$. The term containing $\sigma^1$ vanishes identically
for a fixed value of $\theta$, when the following relation involving the
scalar and the pseudo-scalar potential is satisfied:
\be
 P = a_0 M + b_0, \ \ a_0, b_0 \in R, \ \ \theta=tan^{-1} a_0.
\ee
\noindent Thus, the Hamiltonian $H_1$ becomes diagonal in this limit and
depends on only one arbitrary function $M(x)$.

The Hamiltonian $H_1$ can be re-written in a manifestly
supersymmetric form,
\bea
\tilde{H} & := & H_1 -\frac{b_0^2}{{1+a_0^2}}\nonumber \\
& = & p_x^2 + W^2 - \sigma_3 W^{\prime},
\eea
\noindent where the superpotential $W(x)$ is introduced as,
\be
W(x) \equiv \sqrt{1+a_0^2} M + \frac{a_0b_0}{\sqrt{1+a_0^2}}.
\ee
\noindent Thus, the known relation between supersymmetry and the Dirac
equation in ${\cal{H}}_D$\cite{khare,nogami} is extended to the case
of non-Dirac-hermitian quantum systems. There is an exhaustive list of
superpotentials $W(x)$ for which $\tilde{H}$ is exactly solvable\cite{khare}.
The non-Dirac-hermitian relativistic Hamiltonian,
\bea
&& H_{1D} = \sigma^2 p_x + \sigma^1 \left [ e^{\phi} M_+ +
e^{-\phi} M_- \right ]
+\sigma^3 \left [ e^{(\phi-i\frac{\pi}{2})} M_+ + e^{-(\phi-i\frac{\pi}{2})}
M_- \right ]\nonumber \\
&& M_{\pm} \equiv \frac{1 \pm i a_0}{\sqrt{1+a_0^2}} W(x) - \frac{b_0}{1+a_0^2}
\left ( a_0 \mp i \right ),
\label{hami_1D}
\eea
\noindent is exactly solvable for each superpotential $W(x)$ for which
$\tilde{H}$ is exactly solvable. Moreover, a consistent quantum description
of $H_{1D}$, including entirely real spectra and unitary time-evolution is
admissible. An eigen-spinor $\Psi=\pmatrix{{\Psi_1}\cr \\ {\Psi_2}}$
of $H_{1D}$ with energy eigenvalues $E$ is related to the corresponding
eigen-functions $\psi=\pmatrix{{\psi_1}\cr \\ {\psi_2}}$ of $\tilde{H}$ with
energy eigenvalues ${\tilde{E}}_{1D}$ through the relation,
\bea
\Psi = \left ( \rho_{1D}^{-1} U \right ) \psi.
\eea
\noindent The orthonormality of $\Psi$ in ${\cal{H}}_{\eta_{1D}}$ is guaranteed,
if the eigenfunctions $\psi$ are constructed as a complete set of orthonormal
eigenfunctions in ${\cal{H}}_D$. The eigen-values $E$ are determined in
terms of the eigenvalues ${\tilde{E}}$ through the relation,
\be
E= \pm \sqrt{{\tilde{E}} + \frac{b_0^2}{{1+a_0^2}}}.
\ee
\noindent The parameter `$b_0$' can be chosen as zero without any loss of
generality, since ${\tilde{E}}$ being the energy eigen values of
a supersymmetric theory is always semi-positive definite. In this limit,
$E = \pm \sqrt{\tilde{E}}$ and the zero-energy ground state of
the supersymmetric Hamiltonian $\tilde{H}$ corresponds to the
zero modes of the Dirac Hamiltonian $H_{1D}$ given in Eq. (\ref{hami_1D}).
The choice of a reflection-less superpotential  $W(x)$ produces a perfectly
transparent potential for the non-Dirac-hermitian Hamiltonian.

\subsection{Case III: Scalar, Pseudo-scalar \& Vector interactions}
The third example presented in this article is for non-vanishing $M(x)$,
$P(x)$ and vector potential $V(x)$. The eigen-value equation of
$h_{1D}$ with arbitrary $P(x), M(x)$ and $V(x)$ reads,
\be
\pmatrix{ {P(x) + V(x)} & {-i p_x + M(x)}\cr \\
{i p_x + M(x)} & {-P(x) + V(x)}} \pmatrix{{\psi^{(+)}}\cr \\ {\psi^{(-)}}} =
E \pmatrix{{\psi^{(+)}}\cr \\ {\psi^{(-)}}},
\label{ev}
\ee
\noindent which can be decoupled by introducing the functions
$\theta_{\pm}(x)$, $\chi_{\pm}(x)$,
\be
\chi_{\pm}(x) \equiv \frac{\theta_{\pm}^{\prime}(x)}{2 \theta_{\pm}(x)}, \ \
\theta_{\pm}(x) \equiv E \pm P(x) - V(x),
\label{ev1}
\ee
\noindent and redefining $\psi^{(\pm)}$ in terms of $\tilde{\psi}^{(\pm)}$ as,
\be
{\tilde{\psi}}^{(\pm)}(x) \equiv {\theta_{\pm}}^{-\frac{1}{2}} \psi^{(\pm)} (x).
\label{ev2}
\ee
\noindent The decoupled equations have the form of stationary state
Schr$\ddot{o}$dinger equations,
\bea
&& H_{\pm} {\tilde{\psi}}^{(\pm)} = E^2 {\tilde{\psi}}^{(\pm)}, \ \
H_{\pm}:= -\frac{d^2}{dx^2} + V_{\pm}(x), \nonumber \\
&& V_{\pm} \equiv M^2 + P^2 - V^2 + 2 E V + \chi_{\pm}^2 
-\chi_{\pm}^{\prime} \mp \left ( M^{\prime} - M \chi_{\pm} \right ).
\eea
\noindent Unlike the previous two cases, the potentials $V_{\pm}$ appearing
in the effective Schr$\ddot{o}$dinger equations explicitly depend on the energy
$E$ of the Dirac equation. In general, the potentials $V_{\pm}$ can not be
identified as partner superpotentials of a supersymmetric Hamiltonian. Such
a scenario may arise for very specific choices of the interactions, the study
of which is beyond the scope of this article. Nevertheless, either $H_+$ or
$H_-$ can be embedded in a supersymmetric theory and the eigen-value equation
can be solved analytically.

The eigen value
equation of the Hamiltonian $h_{1D}$ can be solved by first identifying an
exactly solvable $H_-$ or $H_+$. The eigenfunction $\tilde{\psi}^{(+)}$ of
$H_+$ can then be determined in terms of the eigenfunction $\tilde{\psi}^{(-)}$ 
of $H_-$ or the vice verse, by using the equations (\ref{ev}), (\ref{ev1}),
and (\ref{ev2}). A choice of the vector potential in terms of $P(x)$ as,
\be
V(x) = P(x) + a_1, \ \ a_1 \neq E,\ \ a_1 \in R,
\ee
\noindent leads to vanishing $\chi_+$. With a further choice of
$P(x)= b_1 M(x), b_1 \in R$, the potential $V_{\pm}$ reads,
\bea
V_+ & = & W^2 - W^{\prime} + a_1 (2 E -a_1) - b_1^2 (a_1 -
E)^2,\nonumber \\
V_- & = & W^2 + W^{\prime} + a_1 (2 -a_1) - b_1^2 (a_1 - 
E)^2 + \left [ \chi_-^2 - \chi_-^{\prime} - M \chi_- \right ],\nonumber \\
W & \equiv & M - b_1 (a_1 - E), \ \
\chi_-= -\frac{2 b_1 W^{\prime}(x)}{(1+2b_1^2)(E-a_1) - 2 b_1 W}.
\eea
\noindent Although $H_+$ and $H_-$ do not become super-partners of each other
for this case, $H_+$ can be embedded in a supersymmetric theory. A
supersymmetric quantum system in ${\cal{H}}_D$ may be introduced in terms
of the operators $A$, $A^{\dagger}$ and the Hamiltonian $H^{(1)}, H^{(2)}$
as follows:
\bea
&& A= i p_x + W(x), \ \ A^{\dagger} = -i p_x + W(x)\nonumber \\
&& H^{(1)} := A^{\dagger} A = -\frac{d^2}{d x^2} + W^2
- W^{\prime},\nonumber \\
&& H^{(2)} := A A^{\dagger} = -\frac{d^2}{d x^2} + W^2 + W^{\prime}. \ \
\label{ev3}
\eea
\noindent The eigenfunctions and the corresponding eigenvalues of $H^{(1)}$
and $H^{(2)}$ are denoted as $\psi_n^{(1)}, e_n^{(1)}$ and
$\psi_n^{(2)}, e_n^{(2)}$, respectively. The Hamiltonian $H_+$ and $H^{(1)}$
differ by a constant and commute with each other. Thus, the eigen-values
$E_n^{\pm}$ and the corresponding eigen functions $\psi_n^{(+)}$ of $h_{1D}$
are determined as,
\be
E_n^{\pm} = a_1 \pm \left ( \frac{e_n^{(1)}}{1+b_1^2} \right )^{\frac{1}{2}},\ \
\psi_n^{(+)} = \sqrt{E_n^{\pm} - a_1} \psi_n^{(1)}.
\label{ev4}
\ee
\noindent The convention to be followed henceforth is that for
positive(negative) energy solutions $E^+(E^-)$ will be taken in the
expressions of $\psi_n^{+}$ and $\psi_n^-$.
It follows from Eqs. (\ref{ev}), (\ref{ev1}), (\ref{ev2}),
(\ref{ev3}) and (\ref{ev4}) that
the eigen functions $\psi_n^{(-)}$ of $h_{1D}$ corresponding to energy
eigen values $E_n^{\pm}$ is determined in terms of $\psi_n^{(1)}$ as,
\bea
\psi_n^{(-)} & = & \frac{1}{\sqrt{E_n^{\pm} -a_1}} \left [ A -
b_1(E_n^{\pm} -a_1) \right ] {\psi}_n^{(1)}\nonumber \\
& = & \frac{1}{\sqrt{E_n^{\pm} -a_1}} \left [ \sqrt{e_n^{(1)}} \psi_{n-1}^{(2)}
- b_1(E_n^{\pm} -a_1) {\psi}_n^{(1)} \right ],
\eea
\noindent where $\psi_0^{(2)}=0=\psi_{-1}^{(2)}$. The Hamiltonian $H^{(1)}$
is exactly solvable for several choices of $W(x)$ corresponding to shape
invariant potentials\cite{khare}. Thus, the Hamiltonian $h_{1D}$ and hence,
$H_{1D}$ are also exactly solvable for the specific forms of $V(x)$, $P(x)$
and $M(x)$ mentioned above for which $W(x)$ produces shape-invariant potentials.

A few comments are in order before the end of this section:\\
(i) A choice $V(x)= - P(x) + a_2, \ a_2 \in R$ leads to vanishing $\chi_-$.
With a further choice of $P(x)=b_2 M(x), b_2 \in R$ and following the
procedure outlined above, the complete eigen value equation can be solved
analytically by embedding $H_-$ in a supersymmetric theory.\\
(ii) The connection between Dirac equation with scalar interaction and
supersymmetry is well known\cite{khare,nogami}. It appears that a
connection between supersymmetry and Dirac equation with scalar and
pseudo-scalar interactions has been established for the first time
in this article. The method outlined in this article for the study of
Dirac equation with most general coupling makes an indirect connection
with supersymmetry for specific reductions of different interactions.
Further investigations along this line may lead to exactly solvable
$h_{1D}$ with more general interactions.

\section{Dirac Equation in $2+1$ dimensions}

In $2+1$ dimensions, a non-dissipative non-Dirac-hermitian relativistic
quantum system is introduced as follows,
\bea
&& H_{2D} = \Sigma^1 \Pi^1 + \Sigma^2 \Pi^2 + \sigma^3 \phi(X^1,X^2) +
V(X^1,X^2),\nonumber \\
&& \pi^a = p^a - A^a(x^1,x^2), \ \ \Pi^a=P^a - A^a(X^1, X^2), \ \ a=1, 2,
\eea
\noindent where $\phi$ is the scalar potential. The interaction $V$ is the
time component of a Lorentz vector, while $A^1$ and $A^2$ are the space
components of the same vector. The similarity operator $\rho_{2D}$ and the
metric $\eta_{2D}$ are defined as,
\be
\rho_{2D} = exp[-\phi(L^3+\frac{\sigma^3}{2})],\ \
\eta_{2D}:= \rho_{2D}^2,
\ee
\noindent which can be obtained from Eqs. (\ref{metric}) and (\ref{similarity})
by choosing $n^1=0=n^2, n^3=1$. A further choice of $L^3=0$ allows to have
Rasbha-type spin-orbit interaction with purely imaginary coupling in the
Hamiltonian $H_{2D}$. The results of Ref. \cite{bhaba}
may be reproduced easily in this special limit. The discussion below is for
the general case of $L^3 \neq 0$.

The Hamiltonian $H_{2D}$ is hermitian in the Hilbert space
${\cal{H}}_{\eta_{2D}}$ that is endowed with the metric $\eta_{2D}$ and can be
mapped to a Dirac-hermitian Hamiltonian,
\bea
h_{2D} & = & \rho_{2d} H \rho_{2D}^{-1}\nonumber \\
& = & \sigma^1 \pi^1 + \sigma^2 \pi^2 + \sigma^3 \phi(x^1,x^2) + 
V(x^1,x^2).
\eea
\noindent An analytically solvable $h_{2D}$ corresponds to an exactly
solvable $H_{2D}$. An example in this regard may be the relativistic
Landau problem. In particular, choosing $\phi=0$, $A^1=0$, $A^2= B x^1$,
$V=E x^1$ and using the Landau gauge, the Hamiltonian $h_{2D}$ describes a
fermion in a crossed uniform electric and magnetic fields with magnitude
$E$ and $B$, respectively. Such a Hamiltonian also appears in the context
of graphene and analytical results are known\cite{shank}.

\section{Dirac Equation in $3+1$ dimensions}

A non-Dirac-hermitian relativistic quantum system in $3+1$ dimensions
may be introduced as follows,
\be
H_{3D} = \vec{\alpha} \cdot \vec{p} + \beta \Phi(X^1,X^2,X^3) +
i \beta \gamma_5 \xi(X^1,X^2,X^3) + V(X^1,X^2,X^3),
\ee
\noindent where the Dirac-representation is used for the matrices
$\vec{\alpha}$, $\beta$ and $\gamma_5$: 
\be
\vec{\alpha} = \pmatrix{{0} & {\vec{\sigma}}\cr \\
{\vec{\sigma}} & {0}}, \ \
\beta = \pmatrix{{I} & {0}\cr \\ {0} & {-I}}, \ \
\gamma_5= \pmatrix{{0} & {I}\cr \\ {I} & {0}}.
\label{dirac_mat}
\ee
\noindent The scalar and the pseudo-scalar potentials are denoted by
$\Phi(X^1,X^2,X^3)$ and $\xi(X^1,X^2,X^3)$, respectively. The potential
term $V$ can be interpreted as the time component of a Lorentz four-vector.
It may be noted that the
non-Dirac-hermiticity of $H_{3D}$ is due to the presence of $X^1, X^2, X^3$,
which are complex functions of their arguments.

Introducing the similarity operator $\rho_{3D}$ and the metric operator
$\eta_{3D}$,
\be
\rho_{3D} = \pmatrix{{\rho} & {0}\cr \\
{0} & {\rho}}, \ \ \ \eta_{3D}:= \rho_{3D}^2,
\ee
\noindent a non-Dirac-hermitian realization of the Dirac matrices may be
obtained through the replacement of $\vec{\alpha}, \beta$ by
$\vec{\alpha}_1, \beta_1$:
\be
\vec{\alpha}_1 := \pmatrix{{0} & {\vec{\Sigma}}\cr \\
{\vec{\Sigma}} & {0}}, \ \ \beta_1 := \beta.
\ee
\noindent The $\Gamma$ matrices in this non-Dirac-hermitian realization have
the form,
\be
\Gamma_0:= \beta, \ \ \Gamma_i:= \beta \alpha_i, \ \
\Gamma_5:= \gamma_5.
\ee
\noindent It may be noted that $\vec{\alpha_1} \cdot \vec{P}= \vec{\alpha}
\cdot \vec{p}$, implying that these two quantities are hermitian in
${\cal{H}}_D$ as well as in ${\cal{H}}_{\eta_{3D}}$. The Hamiltonian can be
mapped to a Dirac-hermitian Hamiltonian $h_{3D}$ as,
\bea
h_{3D} & := & \rho_{3D}^{-1} H_{3D} \rho_{3D}\nonumber \\
& = & \vec{\alpha} \cdot \vec{p} + \beta \Phi(x^1,x^2,x^3) +
i \beta \gamma_5 \xi(x^1,x^2,x^3) + V(x^1,x^2,x^3).
\eea
\noindent The Hamiltonian $H_{3D}$ is hermitian in the Hilbert space
${\cal{H}}_{\eta_{3D}}$ with the modified norm $\langle \langle \cdot | \cdot
\rangle \rangle_{\eta_{3D}}$. Exactly solvable Dirac-hermitian relativistic
quantum systems are very few in $3+1$ dimensions and no attempt has been made in
this article to present and/or analyze a non-Dirac-hermitian quantum system
with such properties.

\section{Summary \& Conclusions}

A class of non-dissipative non-Dirac-hermitian relativistic quantum systems
those are isospectral with known Dirac-hermitian Hamiltonian has been
constructed in one, two and three space dimensions. A non-Dirac-hermitian
realization of the Dirac matrices and the bosonic operators which are
hermitian in a Hilbert space that is endowed with a positive-definite
metric enables to construct this class of relativistic quantum systems.
The fundamental canonical commutation and anti-commutation relations among
different variables remain unchanged for the non-Dirac-hermitian realization.
Exact solvability of a subclass of these systems has been shown by establishing
and/or employing a connection between supersymmetry and Dirac equation. An
indirect connection between Dirac equation with the most general coupling and
supersymmetry has been established in the process. It is to be seen whether or
not any one or more non-Dirac-hermitian Dirac equations from the large class
of models considered in this article have relevance in the realistic physical
world.

.
\end{document}